%% file: main_arxiv.tex
\documentclass[letterpaper,10pt]{article}
\usepackage[utf8]{inputenc} %unicode support
\usepackage{url}
\usepackage{graphicx}
\usepackage{amsmath,amssymb}
\usepackage{booktabs}
\usepackage{wasysym}
\usepackage{xcolor}
\usepackage[numbers]{natbib}
\usepackage[affil-it]{authblk}
\usepackage{hyperref}

\graphicspath{{../figs/}}

\def \bigfig{\columnwidth}
\def \smallfig{\columnwidth}

\title{How algorithmic popularity bias hinders or promotes quality}
\author[a]{Azadeh Nematzadeh \thanks{Corresponding author: azadeh.n@gmail.com}}
\author[b]{Giovanni Luca Ciampaglia} 
\author[a,b]{Filippo Menczer}
\author[a,b]{Alessandro Flammini}

\affil[a]{School of Informatics and Computing, Indiana University, Bloomington}
\affil[b]{Indiana University Network Science Institute}

\date{}

\begin{document}

\maketitle

\input{abstract}

\input{content}

\section*{Acknowledgements}
\input{acknowledgements}
% Bibliography
\bibliographystyle{plain}
\bibliography{ref}

\end{document}

%% file: abstract.tex
\begin{abstract}
Algorithms that favor popular items are used to help us select among many choices, from engaging articles on a social media news feed to songs and books that others have purchased, and from top-raked search engine results to highly-cited scientific papers. The goal of these algorithms is to identify high-quality items such as reliable news, beautiful movies, prestigious information sources, and important discoveries --- in short, high-quality content should rank at the top.  
Prior work has shown that choosing what is popular may amplify random fluctuations and ultimately lead to sub-optimal rankings. Nonetheless, it is often assumed that recommending what is popular will help high-quality content ``bubble up'' in practice. Here we identify the conditions in which popularity may be a viable proxy for quality content by studying a simple model of cultural market endowed with an intrinsic notion of quality. A parameter representing the cognitive cost of exploration controls the critical trade-off between quality and popularity.  We find a regime of intermediate exploration cost where an optimal balance exists, such that choosing what is popular actually promotes high-quality items to the top. Outside of these limits, however, popularity bias is more likely to hinder quality. 
These findings clarify the effects of algorithmic popularity bias on quality outcomes, and may inform the design of more principled mechanisms for techno-social cultural markets.
\end{abstract}

%% file: content.tex
\section*{Introduction}

Cultural markets, such as social media, the entertainment industry, and the world of fashion are known for their continuous rate of innovation and inherent unpredictability. Success of individual actors (e.g., artists) or products (e.g., songs, movies, memes) is in fact hard to predict in these systems~\cite{Weng2012,Lilian2013srep,Park:2016:SAI:2818048.2820065}, mainly due to the presence of strong social reinforcement, information cascades, and the fact that quality is ultimately predicated on intangible or highly subjective notions, such a beauty, novelty, or virality.

In the absence of objective and readily measurable notions of quality, easily accessible metrics of success --- such as the number of downloads of a song, or the number of social media followers of an individual --- are often taken as input for future recommendations to potential consumers. Popularity and engagement metrics are intuitive and scalable proxies for quality in predictive analytics algorithms. As a result, we are exposed daily to rankings that are based at least partially on popularity, from bestseller lists to the results returned by search engines in response to our queries~\cite{Brin98}. 

The usefulness of such rankings is predicated on the \emph{wisdom of the crowd}~\cite{Surowiecki:2005}: high-quality choices will gain early popularity, and in turn become more likely to be selected because they are more visible. Furthermore, knowledge of what is popular can be construed as a form of social influence; an individual's behavior may be guided by choices of peers or neighbors~\cite{muchnik2013social,lerman2014leveraging,lorenz2011social,krumme2012quantifying,salganik2006experimental,bakshy2011everyone,christakis2009connected}. 
These mechanisms imply that, in a system where users have access to popularity or engagement cues (like ratings, number of views, likes, and so on), high-quality content will ``bubble up'' and allow for a more cost-efficient exploration of the space of choices. This is such a widely shared expectation that social media and e-commerce platforms often highlight popular and trending items. 

Popularity based metrics, however, can bias future success in ways that do not reflect or that hinder quality. 
This can happen in different ways. First, lack of independence and social influence among members of the crowd --- as that implicitly induced by the availability of rankings ---  severely undermines the reliability of the popularity signals~\cite{Lorenz31052011}. Second, engagement and popularity metrics are subject to manipulation, for example by fake reviews, social bots, and astroturf~\cite{Truthy_icwsm2011class,socialbots-CACM}. 

Popularity bias can have more subtle effects. The use of popularity in ranking algorithms by search engines was alleged to impede novel content from rising to the top, but such an entrenchment effect was shown to be mitigated by diverse user queries~\cite{Fortunato22082006}. In social media, some memes inevitably achieve viral popularity in the presence of competition among networked agents with limited attention, irrespective of quality~\citep{Weng2012}, and the popularity of memes follows a power-law distribution with very heavy tails~\citep{refgleeson}. Mechanisms such as unfriending and triadic closure facilitate the formation of homogeneous ``echo chambers'' \cite{Sunstein1} or ``filter bubbles'' \cite{Pariser} that may further distort engagement metrics by selective exposure. 

Even in the absence of engineered manipulation or social distortion, quality is not necessarily correlated with popularity. Consumers face a trade-off between performing cognitively expensive but accurate assessments based on quality and cognitively cheaper but less accurate choices based on popularity. 
Adler %\citeauthor{Adler85} 
has shown that the cost of learning about quality will lead to ``stars'' with disproportionate popularity irrespective of differences in quality~\cite{Adler85}. Such trade-offs are common in social learning environments~\cite{Rendell208}.  
Salganik \textit{et al.} %\citeauthor{salganik2006experimental}
created a music-sharing platform to determine under which conditions one can predict popular musical tracks~\cite{salganik2006experimental}. The experiments showed that in the absence of popularity cues, a reliable proxy for quality could be determined by aggregate consumption patterns. However, popularity bias --- for example when users were given cues about previous downloads of each track --- prevented the quality ranking from being recovered. 
By influencing choices, popularity bias can reinforce initial fluctuations and crystallize a ranking that is not necessarily related to the inherent quality of the choices~\cite{hogg2015disentangling}.
This can happen even in the absence of explicit social signals, if the observed ranking is biased by popularity~\cite{hodas2012limited}. 
Similar results have been found in other studies~\cite{salganik2008leading,salganik2009web,krumme2012quantifying,gilbert2013widespread} and have spurred a renewed interesting in the topic of predictability in cultural markets. 
Van Hentenryck \textit{et al.} % \citeauthor{van2016aligning} 
studied a model of trial-offer markets to analyze the effect of social influence in market predictability~\cite{van2016aligning}. In this model, users chose from a list of items ranked by quality rather than popularity; this modification makes the market predictable and aligns popularity and quality.
Becker \textit{et al.} addressed the question of which network structure is most conducive to the wisdom of the crowd when people are influenced by others~\cite{becker2017network}.

The conditions in which popularity bias promotes or hinders quality content have not been systematically explored. Here we do so by studying
an idealized cultural market model in which agents select among competing items, each with a given quality value.  
A parameter regulates the degree to which items are selected on the basis of their popularity rather than quality. 
We find that this popularity bias yields a rich behavior when combined with the cognitive cost of exploring less popular items. There exists an optimal trade-off in which some popularity bias results in maximal average quality, but this trade-off depends on the exploration cost.

\section*{Results}

Our model considers a fixed number $N$ of items. These represent transmissible units of information, sometimes referred to as \emph{memes}~\cite{dawkins1989selfish}, such as music tracks, videos, books, fashion products, or links to news articles. Items are selected sequentially at discrete times. Each item $i$ has an \emph{intrinsic quality} value $q_i$ drawn uniformly at random from $[0,1]$. Quality is operationally defined as the probability that an item is selected by a user when not exposed to the popularity of the item. The \emph{popularity} of item $i$ at time $t$, $p_i(t)$, is simply the number of times $i$ has been selected until $t$. At the beginning each item is equally popular: $p_i(0)=1,~i=1\ldots N$.

At each time step, with probability $\beta$, an item is selected based on its popularity. All items are first ranked by their popularity, and then an item is drawn with probability proportional to its rank raised to some power:
\begin{equation}
P_i(t) = \frac{r_i(t)^{-\alpha}}{\sum_{i=1}^N r_i(t)^{-\alpha}}
\label{eq:rankprob}
\end{equation}
where the rank $r_i(t)$ is the number of items that, at time $t$, have been selected at least as many times as $i$. The exponent $\alpha$ regulates the decay of selection probability for lower-ranked items. This schema is inspired by the \emph{ranking model}, which allows for the emergence of scale-free popularity distributions with arbitrary power-law exponents~\cite{fortunato2006scale}; it is consistent with empirical data about how people click search engine results~\cite{Fortunato22082006} and scroll through social media feeds~\cite{QiuNHB17}. 
This model could accurately capture aggregate  behavior even if individuals followed different selection schemes~\cite{estes1956problem}.

Alternatively, with probability $1-\beta$, an item is drawn with probability proportional to its quality: 
\begin{equation}
P_i = \frac{q_i}{\sum_{i=1}^N q_i}.
\label{eq:qualityprob}
\end{equation}
After an item $i$ has been selected, we update its popularity ($p_i(t+1) = p_i(t) + 1$) and the ranking. 
Two items will have the same rank $r$ if they have been selected the same number of times. If $k$ item are all at the same rank $r$, then the next rank will be $r + k$.

The model has two parameters: $\beta$ regulates the importance of popularity over quality and thus represents the \emph{popularity bias} of the algorithm. When $\beta=0$, choices are entirely driven by quality (no popularity bias). When $\beta=1$, only popularity choices are allowed, yielding a type of Polya urn model~\cite{Mahmoud:2008}.
The parameter $\alpha$ can be thought of as an \emph{exploration cost}. A large $\alpha$ implies that users are likely to consider only one or a few most popular items, whereas a small $\alpha$ allows users to explore less popular choices. In the limit $\alpha \rightarrow 0$, the selection no longer depends on popularity, yielding the uniform probability across the discrete set of $N$ items.

We vary $\beta$ systematically in $[0,1]$ and consider different values of $\alpha$ between 0 and 3. We simulate 1,000 realizations for each parameter configuration. In each realization we perform $T=10^6$ selections using
%either Eq.~\ref{eq:popularityprob} and~\ref{eq:qualityprob} (Model~\#1), or 
Eqs.~\ref{eq:rankprob} and~\ref{eq:qualityprob} 
%(Model~\#2), 
and store the final popularity values. 

We characterize two properties of the final distribution of popularity $\left\{p_i\right\}_{i=1}^N$ with respect to the intrinsic quality distribution $\left\{q_i\right\}_{i=1}^N$. For brevity, we pose $p_i=p_i(T)$ here.
The first quantity we measure is the \emph{average quality} 
%
%\begin{equation}
%$\bar{q} = \frac{\sum_{i=1}^N p_i q_i}{\sum_{i=1}^N p_i}$ 
$\bar{q} = \sum_{i=1}^N p_i q_i / \sum_{i=1}^N p_i$ 
%\end{equation}
%
and the second property $\tau$ is the \emph{faithfulness} of the algorithm, i.e., the degree to which quality is reflected in  popularity. We quantify faithfulness using Kendall's rank correlation between popularity and quality~\cite{kendall1990rank}.
We can derive the values of both properties in the extreme cases of maximum or no popularity bias. When $\beta=0$, selections are made exclusively on the basis of quality and therefore one expects $p_i \rightarrow q_i$ as $T \rightarrow \infty$. The rankings by quality and popularity are therefore perfectly aligned, and $\tau = 1$.  In the limit of large $N$ we can make a continuous approximation $\bar{q} \rightarrow \int_0^1 q^2 dq / \int_0^1 q dq = 2/3$. When $\beta=1$, quality never enters the picture and any permutation of the items is an equally likely popularity ranking, which translates into $\tau = 0$. Also $p_i \rightarrow 1/N$ and in the continuous approximation $\bar{q} \rightarrow \int_0^1 q dq = 1/2$. What happens for intermediate values of popularity bias is harder to predict.
The question we ask is whether it is possible to leverage some popularity bias to obtain a higher average quality, even at the cost of decreasing the algorithm's faithfulness. 
%%%%% SINCE SANDRO DID NOT RESPOND, LET US ASSUME HE'S SATISFIED WITH THIS CORRECTION.
%\fil{I THINK THE FOLLOWING SENTENCE CAN BE CUT AND THIS WILL ADDRESS SANDRO'S CONCERN. In other words, can an algorithm use popularity bias to take advantage of the ``wisdom of the crowd,'' so that higher-quality items are consumed on average?}

\begin{figure*}[t] 
\centering 
\includegraphics[width=\bigfig]{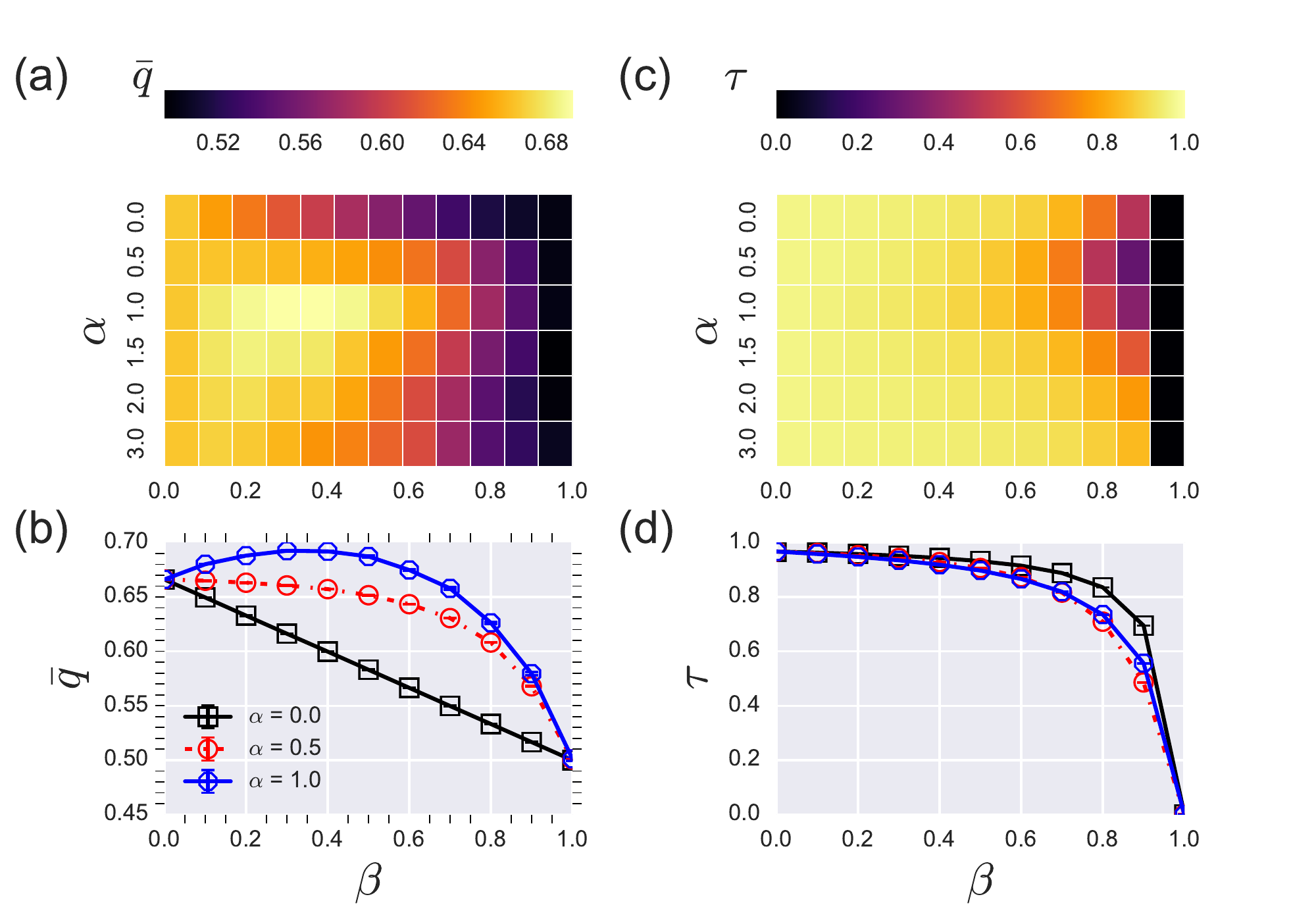}
\caption{\textbf{Effects of popularity bias on average quality and faithfulness.}. 
%Unlike Model~\#1, a small degree of reinforcement maximizes the average quality $\bar q$. 
(\emph{a})~Heatmap of average quality $\bar{q}$ as a function of $\alpha$ and $\beta$, 
% FIL SAYS: I ASSUME THIS IS MODEL 2, NOT 1
%in Model~\#1, 
showing that $\bar{q}$ reaches a maximum for $\alpha=1$ and $\beta \approx 0.4$, while for $\alpha=3$ the maximum is attained for a lower $\beta$. (\emph{b})~The location of the maximum $\bar q$ as a function of $\beta$ depends on $\alpha$, here shown for $\alpha = 0, 0.5, 1.0$. (\emph{c})~Faithfulness $\tau$ of the algorithm as a function of $\alpha$ and $\beta$. (\emph{d})~$\tau$ as a function of $\beta$ for the same three values of $\alpha$. Standard errors are shown in panels (\emph{b,d}) and are smaller than the markers.
}
\label{fig:model2}
\end{figure*}

The dependence of the average quality $\bar q$ on the popularity bias $\beta$ and exploration cost $\alpha$ is shown in Fig.~\ref{fig:model2}(a,b).
We observe that if $\alpha$ is small, popularity bias only hinders quality; the best average quality is obtained for $\beta=0$. However, if $\alpha$ is sufficiently large, an optimal value of $\bar q$ is attained for $\beta>0$. The location of the maximum, $\hat{\beta}=\arg\max_\beta \bar{q}(\beta)$, depends non-trivially on the exploration cost $\alpha$. When popularity-based choices are strongly focused on the top-ranked items ($\alpha > 1$), $\hat\beta$ is a decreasing function of  $\alpha$. Overall, the highest value of $\bar{q}$ is observed for  $\alpha = 1$ and $\beta \approx 0.4$.

In Fig.~\ref{fig:model2}(c,d) we show the behavior of faithfulness $\tau$ as a function of $\alpha$ and $\beta$. We observe that popularity bias always hinders the algorithm's faithfulness, however the effect is small for small $\beta$. This suggests that in the regime where popularity bias improves quality on average, there is a small price to be paid in terms of over-represented low-quality items and under-represented higher-quality items. If these mistakes occur in the low-quality range, they will not affect the average quality significantly. In general, the algorithm can retain faithfulness in the presence of moderate popularity bias, even when the average quality is poor.  
When $\alpha$ is large, $\tau$ remains high over a wide range of popularity bias values. 
In this regime, the preference for popular items is so strong that the vast majority of items (those that do not make the top of the ranking early on) are chosen only by quality-based choice, and therefore their relative ranking perfectly reflects quality. The average quality is however hindered by the top-ranked items, which are selected via popularity irrespective of low quality.

\begin{figure} 
\centering
\includegraphics[width=\smallfig]{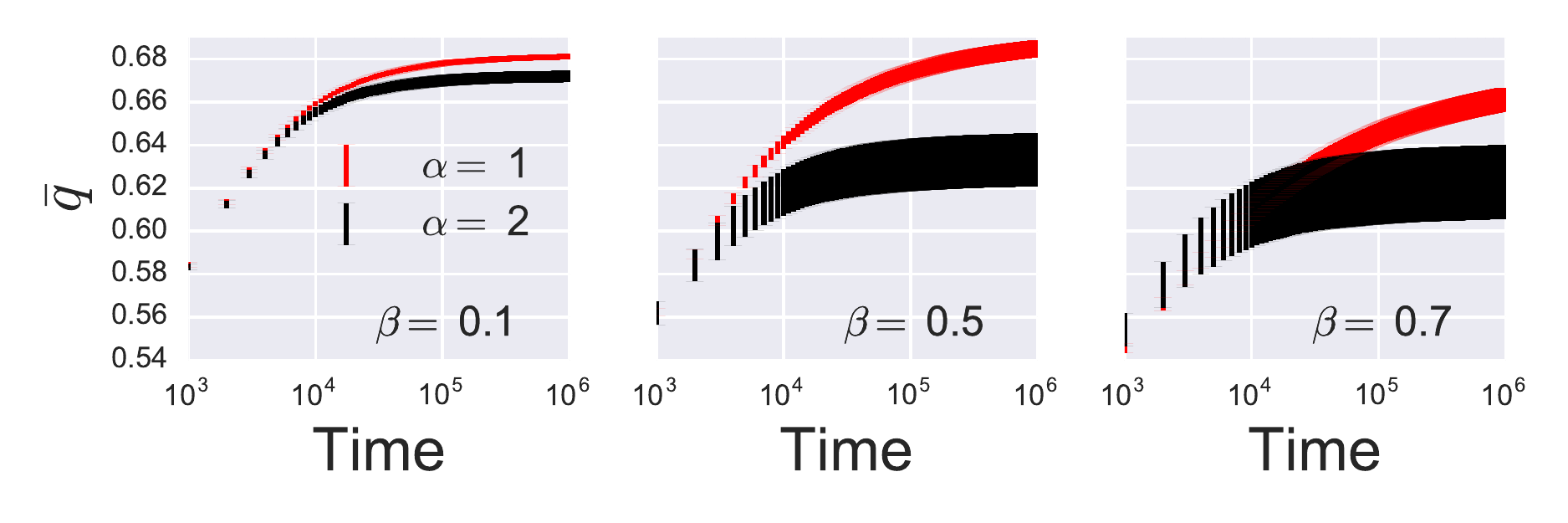}
\caption{\textbf{Temporal evolution of average quality}. Average quality $\bar{q}$ is traced over time for different values of popularity bias $\beta$, in two cases of higher and lower exploration ($\alpha=1$ and $\alpha=2$, respectively). Error bars represent standard errors across runs. With less exploration the system converges early to sub-optimal quality.}% \glc{Use \texttt{tight\_layout} and reduce the number of points (for example, plot only 1,000, 5,000, 10,000, 50,0000, etc); why do the simulations start at $t=10^3$? Change the xlabel to ``Time'', and not ``log(Time)''}}
\label{fig:time_quality}
\end{figure}

In summary, our results show that some popularity bias, together with a mild exploration cost, can produce excellent average quality with minimal loss in faithfulness. Optimizing the average quality of consumed items requires a careful balancing of quality- and popularity-based choices as well as a fine tuning of the focus on the most popular items. For a given value of $\beta$, if $\alpha$ is too low, the popularity bias hinders quality because it fails to enhance the signal provided by the quality-based choices. To understand why quality is also hindered by the popularity bias when $\alpha$ is too high, consider the evolution of the average quality in simulations of the model for different values of $\alpha$ and $\beta$, shown in Fig.~\ref{fig:time_quality}. 
By focusing only on the top ranked items ($\alpha=2$), the system converges prematurely to a sub-optimal ranking, producing lower quality on average. In other words, with insufficient exploration the popularity bias risks enhancing initial noise rather than the quality-based signal. 
With more exploration ($\alpha=1$), $\bar{q}$ continues to grow. %\glc{It is really hard to see in the figure that the convergence time is slower...}. 
The premature convergence to sub-optimal ranking caused by excessive popularity bias is also reflected in the increased variance of the average quality across runs of the model for larger values of both $\alpha$ and $\beta$. This is consistent with the increase in variance of outcomes observed in other studies~\cite{salganik2006experimental,hogg2015disentangling}.

\section*{Discussion}

Cultural markets like social media and the music and fashion industry account for multi-billion businesses with worldwide social and economic impact~\cite{Park:2016:SAI:2818048.2820065}.  Success in these markets may strongly depend on structural or subjective features, like competition for limited attention~\cite{Weng2012,QiuNHB17}. The inherent quality of cultural products is often difficult to establish, therefore relying on measurable quantitative features like the popularity of an item is hugely advantageous in terms of cognitive processing and scalability. 

Yet, previous literature has shown that recommending already popular choices can be detrimental to the predictability and overall quality of a cultural  market~\cite{salganik2006experimental}. This left open the question of whether there exist situations in which a bit of popularity bias can help high-quality items bubble up in a cultural market. 

In this paper we answered this question using an extremely simplified abstraction of cultural market, in which items are endowed with inherent quality. The model could be extended in many directions, for example assuming a population of networked agents with heterogeneous parameters. However, our approach leads to very general findings about the effects of popularity bias. 
While we confirmed that such a bias distorts assessments of quality, the scenario emerging from our analysis is less dire than suggested by prior literature. First, it is possible to maintain a good correspondence between popularity and quality rankings of consumed items even when our reliance on popularity for our choices is relatively high. Second, it is possible to carefully tune the popularity mechanisms that drive our choices to effectively leverage the wisdom of the crowd and boost the average quality of consumed items. 

From a normative perspective, our results provide a recipe for improving the quality of content in techno-social cultural markets driven by engagement metrics, such as social media platforms. It is possible in these systems to estimate the exponent $\alpha$ empirically, by measuring the probability that a user engages with an item as a function of the item's position in the feed. Given a statistical characterization (e.g., average or distribution) of the exploration cost, the bias $\beta$ of the ranking algorithm can be tuned to maximize expected average quality.

These findings are important because in our information-flooded world we increasingly rely on algorithms to help us make consumption choices. Platforms such as search engines, shopping sites, and mobile news feeds save us time but also bias our choices. Their algorithms are affected by and in turn affect the popularity of products, and ultimately drive what we collectively consume in ways that we do not entirely comprehend. It has been argued, for example, that the engagement bias of social media ranking algorithms is partly responsible for the spread of low-quality content over high-quality material~\cite{TowReport2017}. Evaluating such a claim is challenging, but the present results may lead to a better understanding of algorithmic bias.

%% file: acknowledgements.tex
The authors would like to thank Diego F.M. Oliveira and Kristina Lerman for helpful feedback. This work was supported in part by the James S. McDonnell Foundation  (grant 220020274) and the National Science Foundation (award CCF-1101743).

%% file: main_arxiv.bbl
\begin{thebibliography}{10}

\bibitem{Adler85}
Moshe Adler.
\newblock Stardom and talent.
\newblock {\em American economic review}, 75(1):208--12, 1985.

\bibitem{bakshy2011everyone}
Eytan Bakshy, Jake~M Hofman, Winter~A Mason, and Duncan~J Watts.
\newblock Everyone's an influencer: quantifying influence on twitter.
\newblock In {\em proc. of 4th int. conf. on Web search and data mining}, pages
  65--74, 2011.

\bibitem{becker2017network}
Joshua Becker, Devon Brackbill, and Damon Centola.
\newblock Network dynamics of social influence in the wisdom of crowds.
\newblock {\em Proceedings of the national academy of sciences},
  114(26):E5070--E5076, 2017.

\bibitem{TowReport2017}
Emily~J. Bell, Taylor Owen, Peter~D. Brown, Codi Hauka, and Nushin Rashidian.
\newblock The platform press: how {S}ilicon {V}alley reengineered journalism.
\newblock Technical report, Tow Center for Digital Journalism, 2017.

\bibitem{Brin98}
S.~Brin and L.~Page.
\newblock The anatomy of a large-scale hypertextual web search engine.
\newblock In {\em proc. of 7th int. conf. on World-Wide Web (WWW)}, pages
  107--117, 1998.

\bibitem{christakis2009connected}
Nicholas~A Christakis and James~H Fowler.
\newblock {\em Connected: The surprising power of our social networks and how
  they shape our lives}.
\newblock Little, Brown, 2009.

\bibitem{dawkins1989selfish}
R.~Dawkins.
\newblock {\em The selfish gene}.
\newblock Oxford university press, 1989.

\bibitem{estes1956problem}
William~K Estes.
\newblock The problem of inference from curves based on group data.
\newblock {\em Psychological bulletin}, 53(2):134--140, 1956.

\bibitem{socialbots-CACM}
Emilio Ferrara, Onur Varol, Clayton Davis, Filippo Menczer, and Alessandro
  Flammini.
\newblock The rise of social bots.
\newblock {\em Communication of the ACM}, 59(7):96--104, 2016.

\bibitem{Fortunato22082006}
S.~Fortunato, A.~Flammini, F.~Menczer, and A.~Vespignani.
\newblock Topical interests and the mitigation of search engine bias.
\newblock {\em Proceedings of the national academy of sciences},
  103(34):12684--12689, 2006.

\bibitem{fortunato2006scale}
Santo Fortunato, Alessandro Flammini, and Filippo Menczer.
\newblock Scale-free network growth by ranking.
\newblock {\em Physical review letters}, 96(21):218701, 2006.

\bibitem{gilbert2013widespread}
Eric Gilbert.
\newblock Widespread underprovision on reddit.
\newblock In {\em proc. of the 2013 conference on computer supported
  cooperative work (CSCW)}, pages 803--808, 2013.

\bibitem{refgleeson}
James~P Gleeson, Jonathan~A Ward, Kevin~P O'Sullivan, and William~T Lee.
\newblock Competition-induced criticality in a model of meme popularity.
\newblock {\em Physical review letters}, 112(4):048701, 2014.

\bibitem{van2016aligning}
Pascal~Van Hentenryck, Andrés Abeliuk, Franco Berbeglia, Felipe Maldonado, and
  Gerardo Berbeglia.
\newblock Aligning popularity and quality in online cultural markets.
\newblock In {\em proc. of 10th int. AAAI conf. on Web and social media
  (ICWSM)}, pages 398--407, 2016.

\bibitem{hodas2012limited}
Nathan~O Hodas and Kristina Lerman.
\newblock How limited visibility and divided attention constrain social
  contagion.
\newblock In {\em proc. of ASE/IEEE int. conf. on social computing}, 2012.

\bibitem{hogg2015disentangling}
Tad Hogg and Kristina Lerman.
\newblock Disentangling the effects of social signals.
\newblock {\em Human computation}, 2(2):189--208, 2015.

\bibitem{kendall1990rank}
M.G. Kendall and J.D. Gibbons.
\newblock {\em Rank correlation methods}.
\newblock E. Arnold, 1990.

\bibitem{krumme2012quantifying}
Coco Krumme, Manuel Cebrian, Galen Pickard, and Sandy Pentland.
\newblock Quantifying social influence in an online cultural market.
\newblock {\em PloS one}, 7(5):e33785, 2012.

\bibitem{lerman2014leveraging}
Kristina Lerman and Tad Hogg.
\newblock Leveraging position bias to improve peer recommendation.
\newblock {\em PloS one}, 9(6):e98914, 2014.

\bibitem{lorenz2011social}
Jan Lorenz, Heiko Rauhut, Frank Schweitzer, and Dirk Helbing.
\newblock How social influence can undermine the wisdom of crowd effect.
\newblock {\em Proceedings of the national academy of sciences},
  108(22):9020--9025, 2011.

\bibitem{Lorenz31052011}
Jan Lorenz, Heiko Rauhut, Frank Schweitzer, and Dirk Helbing.
\newblock How social influence can undermine the wisdom of crowd effect.
\newblock {\em Proceedings of the national academy of sciences},
  108(22):9020--9025, 2011.

\bibitem{Mahmoud:2008}
Hosam Mahmoud.
\newblock {\em Polya Urn Models}.
\newblock Chapman \& Hall/CRC, 2008.

\bibitem{muchnik2013social}
Lev Muchnik, Sinan Aral, and Sean~J Taylor.
\newblock Social influence bias: A randomized experiment.
\newblock {\em Science}, 341(6146):647--651, 2013.

\bibitem{Pariser}
Eli Pariser.
\newblock {\em The filter bubble: How the new personalized Web is changing what
  we read and how we think}.
\newblock Penguin, 2011.

\bibitem{Park:2016:SAI:2818048.2820065}
Jaehyuk Park, Giovanni~Luca Ciampaglia, and Emilio Ferrara.
\newblock Style in the age of instagram: predicting success within the fashion
  industry using social media.
\newblock In {\em proc. of 19th conf. on computer-supported cooperative work \&
  social computing (CSCW)}, pages 64--73, 2016.

\bibitem{QiuNHB17}
Xiaoyan Qiu, Diego F.~M.~Oliveira, Alireza Sahami~Shirazi, Alessandro Flammini,
  and Filippo Menczer.
\newblock Limited individual attention and online virality of low-quality
  information.
\newblock {\em Nature human behavior}, 1(0132):s41562--017, 2017.

\bibitem{Truthy_icwsm2011class}
Jacob Ratkiewicz, Michael Conover, Mark Meiss, Bruno Gon\c{c}alves, Alessandro
  Flammini, and Filippo Menczer.
\newblock Detecting and tracking political abuse in social media.
\newblock In {\em proc. of 5th int. AAAI conf. on weblogs and social media
  (ICWSM)}, pages 297--304, 2011.

\bibitem{Rendell208}
L.~Rendell, R.~Boyd, D.~Cownden, M.~Enquist, K.~Eriksson, M.~W. Feldman,
  L.~Fogarty, S.~Ghirlanda, T.~Lillicrap, and K.~N. Laland.
\newblock Why copy others? insights from the social learning strategies
  tournament.
\newblock {\em Science}, 328(5975):208--213, 2010.

\bibitem{salganik2006experimental}
Matthew~J Salganik, Peter~Sheridan Dodds, and Duncan~J Watts.
\newblock Experimental study of inequality and unpredictability in an
  artificial cultural market.
\newblock {\em Science}, 311(5762):854--856, 2006.

\bibitem{salganik2008leading}
Matthew~J Salganik and Duncan~J Watts.
\newblock Leading the herd astray: An experimental study of self-fulfilling
  prophecies in an artificial cultural market.
\newblock {\em Social psychology quarterly}, 71(4):338--355, 2008.

\bibitem{salganik2009web}
Matthew~J Salganik and Duncan~J Watts.
\newblock Web-based experiments for the study of collective social dynamics in
  cultural markets.
\newblock {\em Topics in cognitive science}, 1(3):439--468, 2009.

\bibitem{Sunstein1}
Cass~R Sunstein.
\newblock {\em Republic.com 2.0}.
\newblock Princeton university press, 2009.

\bibitem{Surowiecki:2005}
James Surowiecki.
\newblock {\em The Wisdom of Crowds}.
\newblock Anchor, 2005.

\bibitem{Weng2012}
L.~Weng, A.~Flammini, A.~Vespignani, and F.~Menczer.
\newblock Competition among memes in a world with limited attention.
\newblock {\em Scientific reports}, 2(335):e335, 2012.

\bibitem{Lilian2013srep}
L.~Weng, F.~Menczer, and Y.-Y. Ahn.
\newblock Virality prediction and community structure in social networks.
\newblock {\em Scientific reports}, 3(2522):e2522, 2013.

\end{thebibliography}
